\documentclass[11pt]{article}
\pdfoutput=1

\usepackage{hyperref}

\usepackage[normalem]{ulem} 
\usepackage{cite}
\usepackage{appendix}
\usepackage{epsfig}
\usepackage{amsmath}
\usepackage{amssymb}
\usepackage{bm}
\usepackage{hyperref}
\usepackage{slashed}
\newcommand{\be} {\begin{equation}}
\newcommand{\ee} {\end{equation}}
\newcommand{\ba} {\begin{eqnarray}}
\newcommand{\ea} {\end{eqnarray}}

\newcommand{\no} {\nonumber}
\newcommand{\cL} {\mathcal L}
\newcommand{\cA} {\mathcal A}
\newcommand{\cB} {\mathcal B}

\newcommand{\GeV}{\text{GeV}}
\newcommand{\lsim}{\stackrel{<}{_\sim}}

\newcommand{\Dlr}{\stackrel{\leftrightarrow}{D}}

\newcommand{\cLT} {{ \mathcal L}^{(T)}}
\newcommand{\llq} {\lambda^q}
\newcommand{\lle} {\lambda^\ell}
\newcommand{\gq}{g_q}
\newcommand{\gl}{g_\ell}

\usepackage{color}
\definecolor{darkblue}{cmyk}{1,0.3,0,0.2}
\definecolor{violet}{cmyk}{0,1,0,0.2}
\hypersetup{colorlinks, bookmarksnumbered, citecolor=darkblue, linkcolor=darkblue, pdfstartview=FitH, urlcolor=darkblue, linktocpage}

\newcommand{\arXhref}[1]{\href{http://arxiv.org/abs/#1}{#1}}

\begin{document}

\begin{flushright}
 ZU-TH-16/15  \\
June 2015
\end{flushright}

\thispagestyle{empty}

\bigskip

\begin{center}
\vspace{1.5cm}
    {\Large\bf  On the breaking of Lepton Flavor Universality in $B$ decays} \\[1cm]
   {\bf Admir Greljo$^{a,b}$, Gino Isidori$^{a,c}$, David Marzocca$^a$}    \\[0.5cm]
  {\em $(a)$  Physik-Institut, Universit\"at Z\"urich, CH-8057 Z\"urich, Switzerland}  \\
  {\em $(b)$  Faculty of Science, University of Sarajevo, Zmaja od Bosne 33-35, \\ 71000 Sarajevo, Bosnia and Herzegovina}  \\  
  {\em $(c)$  INFN, Laboratori Nazionali di Frascati, I-00044 Frascati, Italy}  \\[1.0cm]
\end{center}

\centerline{\large\bf Abstract}
\begin{quote}
\indent
In view of recent experimental indications of violations of Lepton Flavor Universality (LFU) in $B$ decays, 
we analyze constraints and implications of LFU interactions, both using an effective theory approach, 
and an explicit dynamical model. 
We show that a simple dynamical model based on a $SU(2)_L$ triplet of massive vector bosons, coupled predominantly to third generation fermions 
(both quarks and leptons), 
can significantly improve the description of present data.
In particular, the model decreases the tension between data and SM predictions concerning:
i)~the breaking of $\tau$--$\mu$ universality in $B\to D^{(*)} \ell \nu$  decays;
ii)~the breaking of $\mu$--$e$ universality in $B \to K \ell^+\ell^-$ decays.
Indirectly, the model might also decrease the discrepancy between exclusive and inclusive determinations 
of $|V_{cb}|$ and $|V_{ub}|$.  The minimal version of the model is in tension with ATLAS and CMS direct 
searches for the new massive vectors (decaying into $\tau^+\tau^-$ pairs), 
but this tension can be decreased with additional non-standard degrees of freedom.
Further predictions of the model both at low- and high-energies, in view of future high-statistics data, are discussed. 
\end{quote}

\newpage 

\tableofcontents

\section{Introduction}
Recent experimental data in $B$ physics  hint toward deviations of Lepton Flavor Universality (LFU) in semi-leptonic decays, 
both in the case of $b\to c$ charged-current transitions, as well as in the case of $b\to s$ neutral currents. 
The statistically most significant results can be summarized as follows: 
\begin{itemize}
\item{}  $3.8\sigma+2.0\sigma$ deviation of $\tau/\ell$ universality ($\ell = \mu,e$) in $b \to c$ transitions, encoded 
by~\cite{Lees:2013uzd,BelleD*,LHCbD*}:
\ba
R^{\tau/\ell}_{D^*} &=&\frac{ \cB(B \to D^* \tau \nu)_{\rm exp}/\cB(B \to D^* \tau \nu)_{\rm SM} }{ \cB(B \to D^* \ell \nu )_{\rm exp}/ \cB(B \to D^* \ell \nu )_{\rm SM} } = 
 1.28 \pm 0.08~, \label{eq:RDexp} 
 \\
R^{\tau/\ell}_{D } &=& \frac{ \cB(B \to D  \tau \nu)_{\rm exp}/\cB(B \to D  \tau \nu)_{\rm SM} }{ \cB(B \to D  \ell \nu )_{\rm exp}/ \cB(B \to D  \ell \nu )_{\rm SM} } = 
 1.37 \pm 0.18~,
\label{eq:RDDexp}
\ea
\item{}  $2.6\sigma$ deviation of $\mu/e$  universality in $b \to s$ transitions~\cite{Aaij:2014ora}:\footnote{The result in Eqs.~(\ref{eq:RDexp}) and (\ref{eq:RDDexp}) 
are obtained using $\cB(B \to D^* \tau \nu)/\cB(B \to D^* \ell \nu )_{\rm exp} =  0.323 \pm 0.021 $  
and $\cB(B \to D  \tau \nu)/\cB(B \to D  \ell \nu )_{\rm exp} = 0.41 \pm 0.05$  
from the average of Babar~\cite{Lees:2013uzd}, 
Belle~\cite{BelleD*}, and LHCb~\cite{LHCbD*}, 
assuming $e/\mu$ universality in  $b\to c \ell \nu$ decays, as indicated by $b\to c \ell\nu$ data~\cite{PDG}
(see Sect.~\ref{sect:chargcurr}), 
together with the theory predictions $\cB(B \to D^* \tau \nu)/\cB(B \to D^* \ell \nu )_{\rm SM} = 0.252\pm 0.003$~\cite{Fajfer:2012vx}
and  $\cB(B \to D  \tau \nu)/\cB(B \to D \ell \nu )_{\rm SM} =  0.31\pm 0.02$~\cite{Becirevic:2012jf}.
The SM expectation of $R^{\mu/e}_K$ is $|(R^{\mu/e}_K)_{\rm SM}-1| < 1\%$~\cite{Hiller:2003js} while, 
by construction,  $R^{\tau/\ell}_{D^*}=R^{\tau/\ell}_{D}=1$ within the SM. }
\be
R^{\mu/e}_K =  \left. \frac{ \cB(B \to K \mu^+ \mu^-)_{\rm exp} }{ \cB(B \to K e^+ e^- )_{\rm exp} } \right|_{q^2\in[1,6]{\rm GeV}} =  0. 745 {}^{+0.090}_{-0.074} \pm 0.036~.
\label{eq:RKexp}
\ee
\end{itemize}
 
In addition to these LFU ratios, whose deviation from unity would clearly signal physics beyond the Standard Model (SM),
$B$-physics data exhibit other tensions with SM expectations in semi-leptonic observables.
Most notably, a $\sim3 \sigma$ deviation from the SM expectation has been reported by LHCb~\cite{Aaij:2013qta} in the so-called $P_5^\prime$ 
differential observable of $B\to K^*\mu^+\mu^-$ decays~\cite{Descotes-Genon:2013vna}. 
Moreover, in charged current transitions there is a long-standing 
$\sim 2.5\sigma$ discrepancy in the determination of both $|V_{cb}|$ and $|V_{ub}|$ from exclusive vs.~inclusive semi-leptonic 
decays~\cite{Amhis:2014hma}.

These deviations from the SM have triggered a series of theoretical speculations
about possible New Physics (NP) interpretations, see in particular Ref.~\cite{Descotes-Genon:2013wba,Altmannshofer:2013foa,
Datta:2013kja,Hiller:2014yaa,Ghosh:2014awa,
 Sierra:2015fma,
 Celis:2015ara,
 Becirevic:2015asa,
 Varzielas:2015iva,
 Crivellin:2015era,
Crivellin:2014zpa,
Crivellin:2015mga,Glashow:2014iga,Bhattacharya:2014wla,Gripaios:2014tna,Alonso:2015sja}. 
Among these recent papers, two particularly interesting observations are:
i)~the proposal of Ref.~\cite{Glashow:2014iga} to explain both $R^{\mu/e}_K$ and the $P_5^\prime$ anomaly by means of 
NP coupled dominantly to the third generation of  
quarks and leptons, with a small non-negligible mixing between third and second generations; ii)~the observation of 
 Ref.~\cite{Bhattacharya:2014wla} that  is natural to  
establish a connection between  $R^{\mu/e}_K$ and  $R^{\tau/\ell}_{D^*}$  if the effective four-fermion
semi-leptonic operators  
are build in terms of left-handed doublets.

Despite this recent progress, a coherent dynamical picture 
explaining all the anomalies has not emerged yet. 
On the one hand, a significantly improved fit of experimental data 
can be obtained with a specific set of four-fermion operators of the type 
$J_{q }  \times J _{\ell}$, where $J_{q } $ and $J_{\ell} $ are 
flavor-non-universal left-handed quark and lepton currents~\cite{Bhattacharya:2014wla,Alonso:2015sja}.
On the other hand, even within an Effective Field Theory (EFT) approach, it is hard to believe that 
this set of effective operators is the only relevant one in explicit NP models. 
In particular, explicit NP models should face the tight constraints on 
four-quark and four-lepton 
operators dictated by meson-antimeson mixing, 
and by the bounds on Lepton  Flavor Violation  (LFV) and  LF non-universality 
in pure leptonic processes. Moreover, the size of the SM modifications in 
Eqs.~(\ref{eq:RDexp})--(\ref{eq:RKexp}) points toward   relatively light  
 new degrees of freedom, that could well be within the reach (or already excluded) 
 by direct searches at the LHC.

In this paper we present an attempt to build a simplified coherent dynamical model able 
to explain, at least in part, these violations of LFU. The guiding principle of our construction 
is the idea that the $J_{q }  \times J_{\ell }$ effective operators are generated by the 
exchange of one set  (or more sets) of massive vector bosons that transform as a $SU(2)_L$ triplet,
and that are coupled to both quark and lepton currents. This hypothesis allows us to establish a connection between 
quark-lepton, quark-quark, and lepton-lepton effective operators. We further assume that the  
flavor structure of the new currents is consistent with an approximate $U(2)_q \times U(2)_\ell$ flavor symmetry  
acting on the first two generations of quarks and leptons,
along the lines of Ref.~\cite{Barbieri:2011ci}.

Under these assumptions we proceed with two main steps: i)~we analyze the low-energy 
constraints (and  the corresponding phenomenological implications) on the complete set of four-fermion operators  generated
within the model; ii)~we discuss the additional constraints due to 
electroweak precisions test and collider searches, following from 
the specific choice of the mediators.

We find that, after taking into account all the existing constraints, the proposed model can still provide a  significantly 
improved fit as far as low-energy observables are concerned. The most serious constraint on the model follows from 
the searches performed by ATLAS and CMS on new heavy neutral states $(Z^\prime)$ decaying into $\tau^+\tau^-$ pairs. 
However, as we will discuss, the tension with direct searches can be decreased with additional non-standard degrees 
of freedom, whose net effect is the enhancement of the $Z^\prime$ decay width and the corresponding 
suppression of the  $Z^\prime \to \tau^+\tau^-$ branching ratio. The tension  can be further reduced in the limit 
where the assumption of narrow resonances ($\Gamma \ll M$), 
that is implicit in all present direct searches, no longer holds. 

\section{The model}

\subsection{Step I: four-fermion operators} 

Our main assumption is that all the non-standard four-fermion interactions can be described by the following effective Lagrangian
\ba
  \Delta \cLT_{4f} &=&  - \frac{1}{2 m_V^2} J_\mu^a J_\mu^a~,
  \label{eq:DeltaL4f}
 \ea
 where  $J_\mu^a$ is a fermion current  transforming as a $SU(2)_L$ triplet, built in terms of SM quarks and lepton fields:
\be
	J^a_\mu =  \gq \llq_{ij} \left(\bar Q_L^i \gamma_\mu T^a Q_L^j\right) 
+ \gl \lle_{ij} \left(\bar L_L^i \gamma_\mu T^a L_L^j\right)~.
\label{eq:LVf}
\ee
Here $\lambda^{q,\ell}$ are Hermitian flavor matrices and, by convention, $\llq_{33}=\lle_{33}=1$.

We define $Q^i_L$ and $L^i_L$ to be the quark and lepton electroweak doublets in the 
flavor basis where down-type quarks and charged-leptons are diagonal. 
We assume an approximate $U(2)_q \times U(2)_\ell$ flavor symmetry, under which the light generations 
of $Q^i_L$  and $L_L^i$ transform as $2_q\times 1_\ell$  and $1_q\times 2_\ell$, respectively, and all other fermions are singlets.
We further assume that the underlying  dynamics responsible for the effective interaction  in Eq.~(\ref{eq:DeltaL4f})
involves, in first approximation, only third generation SM fermions (the left-handed $1_q\times 1_\ell$ fermions).
In this limit, the flavor couplings in Eq.~\eqref{eq:LVf} are $\lambda^{q,\ell}_{ij} = \delta_{i3}\delta_{3j}$. 
The corrections to this limit are expected to be generated by appropriate  $U(2)_q \times U(2)_\ell$ 
breaking spurions, connected to the generation of subleading terms in the Yukawa couplings for the SM light fermions.

In the quark case, the leading $U(2)_q$ breaking spurion is a doublet, whose
flavor structure is unambiguously connected to the CKM matrix ($V$)~\cite{Barbieri:2011ci}.
We can thus expand $\llq_{ij}$ as follows:
\be
\llq_{ij} = \delta_{i3}\delta_{3j} + (\epsilon_1 \delta_{i3} \hat V_{3j} + \epsilon_1^* \hat V^*_{3i} \delta_{3j}) + 
\epsilon_2 (  \hat V^*_{3i} \hat V_{3j})+\ldots ~, \qquad   \hat V_{3j} = V_{3j}-\delta_{3j} V_{3j}~,
\label{eq:lambdaexp}
\ee
with $\epsilon_2=O(\epsilon_1^2)$.
As we will discuss below, low-energy flavor-physics data imply $\epsilon_i \ll 1$.

The breaking  structure in the lepton sector is less clear, given the intrinsic ambiguity in reconstructing 
the lepton Yukawa couplings  under the (natural) assumption that neutrino masses 
are generated by a see-saw mechanism.\footnote{An attempt to build a consistent neutrino mass matrix 
starting from an approximate $U(2)_\ell$ symmetry broken by small   $U(2)_\ell$ doublets has been discussed 
in Ref.~\cite{Blankenburg:2012nx}.} As we will discuss below, low-energy data are
compatible with the hypothesis that  the leading breaking terms in the lepton sector transform as 
doublets of $U(2)_\ell$.

Among the four-fermion operators generated by the model, the ones most relevant to flavor phenomenology are:
\ba
\Delta  \cLT_{\rm c.c.}  &=&  - \frac{ \gq \gl }{2 m_V^2} \left[ (V \llq )_{ij}  \lle_{ab}
\left(\bar u_L^i  \gamma_\mu  d_L^j\right)    \left(\bar \ell_L^a \gamma_\mu   \nu_L^b\right)   +{\rm h.c.} \right]~,  \\
 \Delta \cLT_{\rm FCNC}  &=& 
 - \frac{ \gq \gl }{4 m_V^2}   
 \lle_{ab}   \left[    \llq_{ij}  \left(\bar d_L^i  \gamma_\mu  d_L^j\right)  -
(V  \llq V^\dagger )_{ij}  \left(\bar u_L^i  \gamma_\mu  u_L^j\right)  \right] \left(
  \bar \ell_L^a \gamma_\mu   \ell_L^b  -  \bar \nu_L^a \gamma_\mu   \nu_L^b\right)~, \label{eq-eff-lag} \\
 \Delta \cLT_{\Delta F=2} &=&  - \frac{ \gq^2 }{8 m_V^2}  \left[
 (\llq_{ij})^2  \left(\bar d_L^i  \gamma_\mu  d_L^j\right)^2  + 
(V \llq V^\dagger )^2_{ij}  \left(\bar u_L^i  \gamma_\mu  u_L^j\right)^2  \right]~, \\
 \Delta \cLT_{\rm LFV} &=&  - \frac{ \gl^2  }{8 m_V^2}   \lle_{ab}    \lle_{cd}   
(  \bar \ell_L^a \gamma_\mu   \ell_L^b  ) ( \bar \ell_L^c \gamma_\mu   \ell_L^d ) ~, \\
 \Delta \cLT_{\rm LFU}  &=&   - \frac{ \gl^2  }{8 m_V^2}   (-2 \lle_{ab}    \lle_{cd} + 4 \lle_{ad}    \lle_{cb})   
(  \bar \ell_L^a \gamma_\mu   \ell_L^b  ) ( \bar \nu_L^c \gamma_\mu   \nu_L^d ) ~. 
\ea

\subsection{Step II: simplified dynamical model} 

In order to generate $\Delta \cLT_{4f}$ in a dynamical way, 
we introduce the heavy spin-1 triplet, $V_\mu^a$ ($a = 1,2,3$), following the general simplified Lagrangian proposed in Ref.~\cite{Pappadopulo:2014qza}. By means of this approach we can describe both models in which the new vector is weakly coupled, such as gauge extension of the SM, and strongly coupled models, such as Composite Higgs models. The simplified Lagrangian reads
\be
	\cL_V = -\frac{1}{4} D_{[\mu} V_{\nu ]}^a D^{[\mu} V^{\nu ] a} + \frac{m_V^2}{2} V_\mu^a V^{\mu a} + g_H V_\mu^a (H^\dagger T^a i \Dlr_\mu H) + V^a_\mu J_\mu^a ~,
	\label{eq:dyn_mod}
\ee
where $T^a=\sigma^a/2$, $D_{[\mu} V_{\nu ]}^a = D_\mu V_\nu^a - D_\nu V_\mu^a$ and $D_\mu V_\nu^a = \partial_\mu V_\nu^a + g \epsilon^{abc} W_\mu^b V_\nu^c$.\footnote{With respect to Ref.~\cite{Pappadopulo:2014qza} we dropped interaction terms with two or more insertions of $V_\mu^a$. While such terms can be relevant for double production, they do not contribute to the low-energy effective Lagrangian at the dimension-6 level and are thus largely unconstrained by low-energy data.}

By integrating out at the tree-level the heavy spin-1 triplet and keeping only effective operators of dimension $\leq 6$, 
we obtain the effective Lagrangian
\be
	\cL_{\text{eff}}^{d = 6} = - \frac{1}{2 m_V^2} J_\mu^a J_\mu^a
	 -\frac{g_H^2}{2 m_V^2} (H^\dagger T^a i \Dlr_\mu H) (H^\dagger T^a i \Dlr_\mu H)
		- \frac{g_H}{m_V^2} (H^\dagger T^a i \Dlr_\mu H) J_\mu^a~.
	\label{eq:EffLagr}
\ee
By construction, the first term is $\Delta \cLT_{4f}$ in Eq.~(\ref{eq:DeltaL4f}). The second term, in the unitary gauge, is simply
\be
	-\frac{g_H^2 v^2}{4 m_V^2} \left( m_W^2 W_\mu^+ W_\mu^- + \frac{m_Z^2}{2} Z_\mu Z_\mu \right) \left( 1 + \frac{h}{v} \right)^4~.
\ee
This term induces an unphysical (custodially-invariant) shift in the $W$- and $Z$-boson masses,\footnote{Within the full model of 
Eq.~\eqref{eq:dyn_mod} this corresponds to a mass mixing between the SM EW gauge bosons and the heavy vector triplet. The relative shift in the heavy vector masses $m_V$ is only of $O( g_H^2 m_W^2 v^2/ m_V^4)$}. 
that can be reabsorbed by a redefinition of $v$, and deviations in the Higgs interactions to $W$ and $Z$ bosons.
The latter are well within the existing bounds for the relevant set of parameters. 
The last term, instead, describes non-universal deviations in the $Z$ and $W$ couplings to SM quarks and leptons
that lead to non-trivial constraints on the parameter space of the model (see sect.~\ref{sect:LEP}).

\section{Low-energy implications of the four-fermion operators}

\subsection{New physics effects in charged currents}
\label{sect:chargcurr}

Since the new interactions are purely left-handed, in the case of charged currents their effect is simply an overall 
(flavor non-universal) rescaling of the SM amplitudes: 
\be
R^{b\to c}_{\ell^i}  =  \frac{ \cA( b\to c ~\ell^i \bar \nu^i )_{\rm SM +NP} }{  \cA( b\to c ~\ell^i \bar \nu^i )_{\rm SM} }
= 1 + R_0 \lle_{ii} 
\left( 1+ \frac{  V_{cs} (\llq_{bs})^* +  V_{cd} (\llq_{bs})^* }{ V_{cb}  } \right)~, 
\ee
where 
\be
R_0  =  \frac{ \gq \gl  m_W^2 }{ g^ 2 m_V^2} \equiv  \frac{ G_F^{(T)} }{ G_F^{\rm SM} }~.
\ee
Using this expression, the LFU breaking ratio in~Eq.~(\ref{eq:RDexp}) assumes the form 
\be
R^{\tau/\ell}_{D^*} \approx  1  + 2 R_0 \; \text{Re}\! \left[ \left( 1  - \frac{\lle_{\mu\mu}+\lle_{ee}}{2}
 \right) 
\left( 1+ \frac{  V_{cs} (\llq_{bs})^* +  V_{cd} (\llq_{bd})^* }{ V_{cb}  } \right) \right] \approx  1+ 2 R_0~,
\label{eq:R0}
\ee
where we have assumed $|\lle_{\mu\mu,ee} | \ll 1$ and $ | \llq_{ij} |  \ll | V_{3i}^* V_{3j} |$. The 
first condition is required by the smallness of deviations from the SM in $b \to c \ell \nu$ decays (see below),
the second condition follows by the consistency of the bounds  from $\Delta F=2$ amplitudes
(see Sect.~\ref{sect:DF2}). 
We are thus able to fix the overall strength of the new effective charged-current interaction (compared to the Fermi coupling):
\be
	R_0 = \frac{1}{2} \left(  R^{\tau/\ell}_{D^*}   - 1 \right) =  0.14 \pm 0.04~.   
	\label{eq:BoundR0}
\ee 
The model predicts the same violation of $\tau/\ell$ universality for all type of $b\to c$ and $b\to u$ transitions.
This implies, in particular,  $R^{\tau/\ell}_{D}= R^{\tau/\ell}_{D^*}$, that is perfectly consistent 
with the experimental result in~Eq.~(\ref{eq:RDDexp}).

In principle, violations of LFU universality are expected also between $b\to c(u) \mu\nu$ and $b\to c(u) e\nu$ modes.
Unfortunately, it is very difficult to get experimental bounds on the latter using published data,
since most of the high-statistics semi-leptonic analyses 
are performed combining $\mu$ and $e$ modes~\cite{PDG}.
Using the PDG fit for the combined $B^\pm/B^0$ sample~\cite{PDG}, 
that is separated for $\mu$ and $e$ modes, we deduce that deviations between 
$\Gamma(b\to c(u) \mu\nu)$ and $\Gamma(b\to c(u) e\nu)$ as large as $\sim 2\%$ are allowed by present data.
Within our model, we expect
\be
\frac{ \Gamma( b\to c(u) ~\mu \bar \nu )_{\rm SM +NP} }{  \Gamma( b\to c(u) ~e \bar \nu )_{\rm SM+NP} }
\approx  1  + 2 R_0  \left(\lle_{\mu\mu}-\lle_{ee}  \right)~.  
\label{eq:bulnu}
\ee
The strong constraints on LFU involving only quarks and leptons of the first two generations 
($\pi$ and $K$ decays, CKM unitarity, and $\mu$ decay~\cite{Antonelli:2010yf}) 
implies $|\lle_{ee}| \ll |\lle_{\mu\mu}|$.
As a result,
the constraints on $\mu$-$e$ charged-current LFU can be used to set the approximate bound
\be
|\lle_{\mu\mu}| \lsim 0.07 \left( \frac{0.15}{R_0} \right)~,
\label{eq:bmue}
\ee
that justifies having neglected $\lle_{\mu\mu}$ in Eq.~(\ref{eq:R0}).
After this bound is imposed, violations of LFU universality in $K$ and $\pi$ semileptonic decays 
turn out to be unobservables, given the additional suppression factor $|V_{ub} (\llq_{bq})^* / V_{uq} |$,
for $q=s,d$, compared to Eq.~(\ref{eq:bulnu}).

The universal $30\%$ excess to $\tau$ semi-leptonic charged-current decays is likely to explain, at least in part, 
the tension between exclusive and inclusive determinations of $|V_{cb}|$ and $|V_{ub}|$.
The argument goes as follows: the $B\to X_{c,u} \tau \nu$ decays followed by $\tau \to X \ell \nu \nu$ represent 
a background for the inclusive $B\to X_{c,u} \ell \nu$ analyses. At present, this background is subtracted 
via montecarlo simulations that assume a SM-like $\cB(B\to X_{c,u} \tau \nu)$~\cite{Nico}. This procedure therefore
{\em underestimates} the background events and leads to an enhanced  $B\to X_{c,u} \ell \nu$ signal.
On the other hand, the problem is not present in the exclusive decays of the type $B\to D^{(*)}\ell\nu$, where the kinematical 
closure of the events prevents the contamination from $\tau$ decays. A precise estimate of the effect would require a re-analysis 
of $B\to X_{c,u} \ell \nu$  data and is beyond the scope of the present paper. However, we note that this effect necessarily 
goes in the right direction (enhanced signal in inclusive modes), and that is likely to be larger in $b\to u$ compared to 
$b\to c$, given the different kinematical cuts. We are then led to the conclusion that the most reliable estimates 
of  $|V_{cb}|$ and $|V_{ub}|$ are those obtained by means of exclusive decays and, more specifically, 
exclusive decays into electron final states.

As a result of this discussion, we urge the experimental collaborations to reanalyze all semi-leptonic charged-current 
$B$ decays {\em without} imposing LFU, both as far as signal and as far as background are concerned.

\subsection{Bounds from $\Delta F=2$}
\label{sect:DF2}

Also in the case of $\Delta F=2$ transitions the new interaction amounts to an overall 
flavor non-universal rescaling of the SM amplitudes. It is therefore convenient to define the 
ratios
\be
R^{\Delta F=2}_{B_q} = \frac{ \cA  ( B_q \to \bar B_q  )_{\rm SM +NP} }{  \cA (B_q \to \bar B_q  )_{\rm SM}  }
= 1 + R_0 \frac{ \gq }{\gl} \frac{ (\llq_{bq})^2   }{  (V_{tb}^* V_{tq})^2  } \times (R^{\rm loop}_{\rm SM})^{-1}~,  
\ee
where\footnote{For the SM amplitude and the definition for the loop function $S_0(x_t)=S_0(m_t^2/m_W^2) \approx 2.4$ see e.g.~Ref~\cite{Buchalla:1995vs}.}
\be  
 R^{\rm loop}_{\rm SM} = \frac{ \alpha_{\rm em} S_0(x_t) }{ 4 \pi s_W^2 } \approx  6.5 \times 10^{-3}~.
\ee
The consistency  with experimental results on down-type $\Delta F=2$ amplitudes, where no significant deviations from the SM are observed 
(up to the $10\%$-$30\%$ level depending on the specific amplitude)  
implies   $ | \llq_{ij} |  \lsim 10^{-1} | V_{3i}^* V_{3j} |$, for $R_0=0.15$ and $\gl/\gq=O(1)$. As anticipated,
this justifies the expansion on the r.h.s.~of Eq.~(\ref{eq:R0}). 

If the corrections of $\llq_{ij}$ from the leading term are generated by   $U(2)_q$ breaking spurions, 
as proposed in Eq.~(\ref{eq:lambdaexp}), the $R^{\Delta F=2}_{B_q}$ terms should respect the 
$U(2)^3$ prediction~\cite{Barbieri:2011ci}
\be
R^{\Delta F=2}_{B_s} = R^{\Delta F=2}_{B_d}
\ee
that, in turn, implies no corrections to the clean ratio $\Delta M_{B_s}/ \Delta M_{B_d}$. 
Furthermore, if the $\epsilon_1$ parameter in the expansion~(\ref{eq:lambdaexp}) is real,
we expect no corrections to the $CP$-violating phases of $B_s$ and $B_d$ mixing.

To discuss the NP impact in $b\to s\ell^+\ell^-$ decays we need to establish a precise upper bound
on $|\llq_{bs}|$.  According to the $U(2)^3$ fit of meson-antimeson mixing 
of Ref.~\cite{Barbieri:2014tja}, we set  $R^{\Delta F=2}_{B_s} \in [0.8, 1.2]$ 
at 95\%CL, that implies 
\be
|\llq_{bs} | <  |\llq_{bs} |_{\rm max} = 0.093   \left|  V_{ts}  \right|  \left|  \frac{ \gl }{\gq}    \right|^{1/2}
\left(  \frac{ 0.15 }{ R_0 }    \right)^{1/2}~.
\ee
 
Even in the limit of negligible $\llq_{ij}$ for $i\not=3$ or $j\not=3$, a potentially sizable contribution to $\Delta C=2$ is generated by 
CKM mixing, starting from the leading term in Eq.~(\ref{eq:lambdaexp}). This can be written as 
 \ba
\Delta  \cL^{(V)}_{\Delta C=2}  &=&   - \frac{1}{\Lambda_{uc}^2}  \frac{ ( V_{ub} V_{cb}^* )^2 }{| V_{ub} V_{cb}^* |^2  }\left(\bar u_L \gamma_\mu  c_L \right)^2 
+{\rm h.c.}~, \\ 
\Lambda _{uc} &=&   
\left[ \frac{G_F}{\sqrt{2} }  R_0 \frac{ \gq }{\gl }   | V_{ub} V_{cb}^* |^2 \right]^{-1/2}  \approx  
  6.9  \times 10^{3}~{\rm TeV}  \times  \left|  \frac{ \gl }{\gq}    \right|^{1/2}   \left(  \frac{ 0.15 }{ R_0 }    \right)^{1/2}~.   
  \ea
 Remarkably, for $\gl/\gq = O(1)$  
 this result is  compatible 
with the existing bounds from CP violation in 
$D$-$\bar D$ mixing that require $\Lambda _{uc} > 3 \times 10^{3}~{\rm TeV}$~\cite{Isidori:2013ez}.
For $R_0 = 0.15$ this fixes $|\frac{ \gq }{\gl } | \lesssim 5.4$.

\subsection{Bounds from LFU and LFV in $\tau$ decays, and neutrino physics} 

LFU in $\tau$ decays has been tested at the permil level. Assuming $\lle_{ij}$ is negligible if $i=e$ or $j=e$, 
and imposing $|\llq_{ij} |  \ll | V_{3i}^* V_{3j} |$, such tests can be used to set stringent limits on 
$|\lle_{\mu\mu}|$ and $|\lle_{\tau\mu}|$. Moreover, a strong limit on the product 
$|\lle_{\tau\mu}|  |\lle_{\mu\mu}|$ follows from the upper bound on $\cB(\tau \to 3 \mu)$.

The relevant modified effective Lagrangians are  
\ba
\Delta \cLT_{\rm LFU} &=& - \frac{g_\ell^2}{2 m_V^2} \left[ \left(\lle_{\mu \mu}  - \frac{1}{2} |\lle_{\tau \mu}|^2 \right) (  \bar \tau_L \gamma_\mu   \mu_L ) ( \bar \nu_\mu  \gamma_\mu  \nu_\tau) + \frac{1}{2} \lle_{\tau \mu} (  \bar \tau_L \gamma_\mu   \mu_L   ) ( \bar \nu_{\tau}  \gamma_\mu  \nu_\tau) + {\rm h.c.} \right]~, \nonumber \\
\Delta \cLT_{\rm LFV}  &=&  -  \frac{1}{\Lambda_{\tau\mu}^2}      
(  \bar \tau_L \gamma_\mu   \mu_L   ) ( \bar \mu_L  \gamma_\mu  \mu_L ) ~, \qquad 
 \Lambda_{\tau\mu} ~=~  \left[ \frac{ G_F}{\sqrt{2}}  R_0 \frac{ \gl  }{\gq}  \lle_{\mu \mu}    \lle_{\tau \mu} \right]^{-1/2}~.
 \ea
As far as LFU tests are concerned, the observable we consider is~\cite{Stugu:1998jv}
\be
	\frac{ \cB(\tau \to \mu \bar\nu \nu ) f(x_e^2)}{\cB(\tau \to e \bar\nu \nu ) f(x_\mu^2)} = \left| 1 + R_0 \frac{g_\ell}{g_q}\left(\lle_{\mu \mu} - \frac{1}{2} |\lle_{\tau \mu}|^2 \right) \right|^2 + \left| \frac{R_0}{2} \frac{g_\ell}{g_q} \lle_{\tau\mu} \right|^2
	= 1.0040 \pm 0.0032~,
	\label{eq:tauLFU}
\ee
where $x_\ell = \frac{m_\ell}{m_\tau}$, $f(x_\mu^2) / f(x_e^2) = 0.9726$ is a phase space factor,  and we summed over neutrinos of 
arbitrary flavor. The numerical result on the r.h.s.~of Eq.~(\ref{eq:tauLFU}) 
is obtained using PDG data~\cite{PDG}.  
Expanding to first order in $R_0$ and assuming  $|\lle_{\tau \mu}|^2 \ll |\lle_{\mu \mu}|$  we obtain
\be 
 \lle_{\mu \mu}   = ( 0.013 \pm  0.011)  \times  \frac{g_q}{g_\ell} \left(  \frac{ 0.15 }{ R_0 }    \right)~.
   \label{eq:LFU_bound}
\ee
This constraint is significantly stronger than the bound from $b\to c \mu(e) \nu$ universality (Eq.~(\ref{eq:bmue})), unless $g_q/g_\ell \gg 1$.

In principle, an independent bound on $|\lambda^\ell_{\mu\mu}|$ can be obtained by neutrino trident production, 
namely muon pair production from $\nu_\mu$ scattering on a heavy nuclei. 
The inclusive cross section $\sigma(\nu_\mu N \to \nu_\mu N \mu^+ \mu^- )$ is proportional to the combination of 
effective couplings $(C_V^2+C_A^2)$ where $C_{V,A}=C_{V,A}^{\textrm{SM}}+\Delta C_{V,A}$~\cite{Altmannshofer:2014pba}, and the 
SM reference values are $C_{V}^{\textrm{SM}}=1/2+2 \sin^2 \theta_W$ and $C_{A}^{\textrm{SM}}=1/2$~\cite{Altmannshofer:2014pba}. The corrections to these couplings in our model are
\be
\Delta C_V = \Delta C_A = \frac{m_W^2 g_\ell^2}{2 m_V^2 g^2} |\lambda^\ell_{\mu \mu}|^2 ~. \qquad
\ee
Combining the reported cross section measurements from the CHARM-II collaboration ($\sigma/\sigma^{\textrm{SM}}=1.58\pm0.57$~\cite{Geiregat:1990gz}) and the CCFR collaboration ($\sigma/\sigma^{\textrm{SM}}=0.82\pm0.28$~\cite{Mishra:1991bv}), we find 
\be
|\lambda^\ell_{\mu \mu}|  < 1.5 \left| \frac{g_q}{g_\ell} \right|^{1/2} \left(  \frac{ 0.15 }{ R_0 }    \right)^{1/2}~,
\ee
that is well below the LFU bound in Eq.~(\ref{eq:LFU_bound}).

As far as  LFV  is concerned, the $\cB(\tau \to 3 \mu)< 2.1 \times 10^{-8}$ bound~\cite{PDG}
 implies $ \Lambda_{\tau\mu}  > 11$~TeV, that can be translated into  
\be
     | \lle_{\mu\mu} \lle_{\tau\mu } | <  0.005 \left|  \frac{g_q}{g_\ell}  \right|     \left(  \frac{ 0.15 }{ R_0 }    \right)~.
     \label{eq:LFVbound}
\ee 
If $|\lle_{\mu\mu}|$ assumes the maximal allowed by  Eq.~(\ref{eq:LFU_bound}), we are left with the bound 
$|\lle_{\tau\mu} | \lsim 0.15$. The latter is compatible with the hypothesis that $\lle_{ij}$ admits an 
expansion similar to that of $\llq_{ij}$ in Eq.~(\ref{eq:lambdaexp}), or that the leading breaking of the $U(2)_\ell$ flavor symmetry 
is determined by spurions transforming as $U(2)_\ell$ doublets.

\subsection{New-physics effects in $b\to s \ell^+\ell^-$} 
\label{sect:FCNCs}

The effective Lagrangian encoding NP effects in  $b\to s \ell^+\ell^- (\ell=e,\mu,\tau)$ is
\ba
 \Delta \cL^{ (V) }_{ b\to s\ell^+\ell^-}  =
 - \frac{2 G_F}{\sqrt{2} } R_0 \llq_{bs} \left(\bar b_L \gamma_\mu  s_L \right) \left(    
  \bar \tau_L \gamma_\mu  \tau_L + \lle_{\mu\mu}  \bar \mu_L \gamma_\mu \mu_L + \lle_{ee}  \bar e_L \gamma_\mu e_L    \right)~. 
  \ea
Using $\Delta \cL^{ (V) }_{ b\to s\ell^+\ell^-}$  to determine modified matching conditions for the Wilson coefficients of the 
most general $b \to s \ell^+ \ell^-$ effective Hamiltonian,
\be
\mathcal{H}^{b \to s}_{\textrm{eff}} = -\frac{4 G_F}{\sqrt{2}} V_{tb} V_{ts}^* \frac{e^2}{16\pi^2} \sum_i (C^\ell_i O^\ell_i + C^{\ell\prime}
_i O^{\ell\prime})+ \textrm{h.c.}~
\ee
leads to
\be
\Delta C^\tau_9 = - 
\Delta C^\tau_{10} = -  \frac{ \pi  R_0 }{\alpha_{\rm em}}  \frac{\llq_{bs}}{V_{tb}^* V_{ts} }~, \qquad 
\Delta C^{\mu(e)}_9 = - \Delta C^{\mu(e)}_{10} =  - \lle_{\mu\mu(ee)} \Delta C^\tau_{10}~,
\label{eq:DC9T}
\ee
where 
\be 
O^\ell_9 = (\bar s_L \gamma^\nu b_L) \bar \ell \gamma_\nu \ell~, \qquad O^\ell_{10} = (\bar s_L \gamma^\nu b_L) \bar \ell 
\gamma_\nu \gamma^5 \ell~.
\ee

Present $b\to s \mu^+\mu^-$ anomalies and $R^{\mu/e}_K$  seems to indicate a 
LF non-universal modification in the Wilson coefficients $C^\mu_{9}$ compared 
to the SM (see e.g.~Ref.~\cite{Descotes-Genon:2013wba,Altmannshofer:2013foa}).
However, a good fit to present data is also obtained assuming 
$\Delta C^{\mu}_9 =  -  \Delta C^{\mu}_{10} \not =0$ and  $\Delta C_{9,10}^e = 0$, 
that is compatible with the modification expected in our NP framework
for $|\lle_{\mu\mu}| \gg |\lle_{ee}|$. The best fit thus obtained implies 
$\Delta C_9^{\mu}=-  \Delta C^{\mu}_{10}= -0.53\pm0.18$~\cite{Altmannshofer:2014rta}.

In order to reproduce  this result within our model   we must impose  
\be 
\llq_{bs} \lle_{\mu\mu } = (3.4 \pm 1.1) \times 10^{-4}~ \times  \left(\frac{0.15}{R_0} \right)~.
\label{eq:mm2}
\ee
This result is in some tension with the bounds on $|\lle_{\mu\mu}|$  and $|\llq_{bs}|$ dictated by 
LFU in $\tau$ decays and $\Delta m_{B_q}$ mixing, respectively. 
To express this tension more clearly, 
it is convenient to normalize Eq.~(\ref{eq:mm2}) to the maximal value
of  $|\llq_{bs}|$  allowed by $\Delta m_{B_q}$ mixing. This leads to 
\be 
\frac{ \llq_{bs} }{ |\llq_{bs}|_{\rm max} }  \left(\frac{R_0}{0.15}\right)^{1/2}  \left|  \frac{ \gl }{\gq}    \right|^{1/2}  \lle_{\mu\mu } 
=   (0.09 \pm 0.03)~,
\label{eq:mm}
\ee
that should be compared with the constraint  on $|\lle_{\mu\mu}|$ from Eq.~(\ref{eq:LFU_bound}).
Given the different scaling  of  Eq.~(\ref{eq:mm}) and Eq.~(\ref{eq:LFU_bound}) in terms of $\gl/\gq$, 
the tension decreases for $|\gl/\gq |  < 1$.  

As far as $b\to s \tau^+\tau^-$ decays are concerned, for $R_0=0.15$ and $\gq=\gl$, we find 
\be
\Delta C^\tau_9 = - \Delta C^\tau_{10} \approx  - 5.6  \times \frac{ \llq_{bs} }{ |\llq_{bs}|_{\rm max} }~, \qquad  {\rm vs.} \qquad 
(C^\tau_9)_{\rm SM} \approx - (C^\tau_{10})_{\rm SM} \approx  4.2~.
\ee
Thus if $\llq_{bs}$ is close to  $|\llq_{bs}|_{\rm max}$, as favored by 
$b\to s \mu^+\mu^-$ anomalies and $R^{\mu/e}_K$, depending on ${\rm arg}(\llq_{bs})$ we have two very different 
non-standard predictions for  $b\to s \tau^+\tau^-$ decays. In the case of maximal constructive interference of NP and SM 
amplitudes, $b\to s \tau^+\tau^-$ rates could be enhanced up to a factor $\approx 5$ over the  SM;  in the case of maximal 
destructive interference, $b\to s \tau^+\tau^-$ rates could  be strongly suppressed (even less than 1/10) compared to the SM expectation.
This possible enhancement or suppression would hold also for the $b\to s \nu_\tau \bar \nu_\tau$ rates, but it would appear ``diluted" 
by a factor of $\approx 3$ in the measurable $b\to s \nu  \bar \nu$  rates summed over all neutrino species.

In principle, the effective four-fermion Lagrangian in Eq.~(\ref{eq:DeltaL4f}) could allow also FCNC--LFV transitions of the 
type $b\to s \ell^\pm_i \ell^\mp_j$, with the largest amplitude  expected for  $b\to s \tau^\pm \mu^\mp$. The latter 
can  be estimated by means of Eq.~(\ref{eq:DC9T}),
with the replacement $\lle_{\mu\mu} \to \lle_{\tau\mu}$.  Given the constraint  on 
$|\lle_{\tau\mu}|$ in Eq.~(\ref{eq:LFVbound}),  
we find that FCNC--LFV  helicity-conserving transitions ($B\to K \tau^\pm \mu^\mp$, 
$B\to K^* \tau^\pm \mu^\mp$, \ldots)  can have rates which are at most 10\% of those of the corresponding di-muon modes in the SM.
Similarly, we find $\cB(B_s \to \tau^\pm \mu^\mp) \lsim 10^{-8}$.
These bounds makes the experimental search of these FCNC--LFV transitions very challenging, at least in the short term. 
We also note that such bounds are saturated only if $\cB(\tau \to 3\mu)$ is just below its current experimental bound.
 
\subsection{Combined fit and discussion}
\label{sect:fit}

\begin{table}[t]
\begin{center}
\begin{tabular}{c|c|c}
	Obs. $\mathcal{O}_i$ & Exp. bound ($\mu_i \pm \sigma_i$) & Def. $\mathcal{O}_i(x_\alpha)$\\ \hline
	\raisebox{-2pt}[2pt][2pt]{$R_0 (D^*)$} & $0.14 \pm 0.04$ & $\epsilon_\ell \epsilon_q$ \\
	\raisebox{-2pt}[2pt][2pt]{$R_0 (D)$} & $0.19 \pm 0.09$ & $\epsilon_\ell \epsilon_q$ \\
	\raisebox{-2pt}[8pt][8pt]{$\Delta R_{b\rightarrow c}^{\mu e}$} & \raisebox{-2pt}[8pt][8pt]{$0.00 \pm 0.01$} & $2 \epsilon_\ell \epsilon_q \lle_{\mu\mu}$ \\
	\raisebox{-2pt}[8pt][8pt]{$\Delta R_{B_s}^{\Delta F = 2}$} & $0.0 \pm 0.1$ & $\epsilon_q^2  |\llq_{bs}|^2 (  |V_{tb}^* V_{ts}|^2 R_{\text{SM}}^\text{loop})^{-1} $ \\
	\raisebox{-2pt}[8pt][8pt]{$\Delta C_9^\mu$} & $-0.53 \pm 0.18$ & $- (\pi/\alpha_{\rm em}) \lle_{\mu\mu} \epsilon_\ell \epsilon_q  \llq_{bs}/  |V_{tb}^* V_{ts}| $ \\
	\raisebox{-2pt}[8pt][8pt]{$\Delta R_{\tau \rightarrow \mu / e}$} & $0.0040 \pm 0.0032$ & $2 \epsilon_\ell^2 \left(\lle_{\mu \mu} - \frac{1}{2} |\lle_{\tau \mu} |^2 \right)$ \\
	\raisebox{-2pt}[8pt][8pt]{$\Lambda_{\tau\mu}^{-2}$} & $(0.0 \pm 4.1) \times 10^{-9\phantom{0}} \, [\GeV^{-2}]$ & $ (G_F/\sqrt{2}) \epsilon_\ell^2 \lle_{\mu \mu} \lle_{\tau \mu}$  \\
	\raisebox{-0pt}[4pt][4pt]{$\Lambda_{uc}^{-2}$} & $ (0.0 \pm 5.6) \times 10^{-14} \, [\GeV^{-2}]$ & $ (G_F/\sqrt{2}) \epsilon^2_q   | V_{ub} V_{cb}^* |^2$	\\
 \end{tabular}
\caption{\label{tab:FlavorFit} Observables entering in the fit with their experimental bound (assuming the uncertainties follow the Gaussian distribution) and the expression in terms of the parameters of our model.}
\end{center}
\end{table}

\begin{figure}[p]
  \centering
  \begin{tabular}{cc}
   \includegraphics[width=0.45\textwidth]{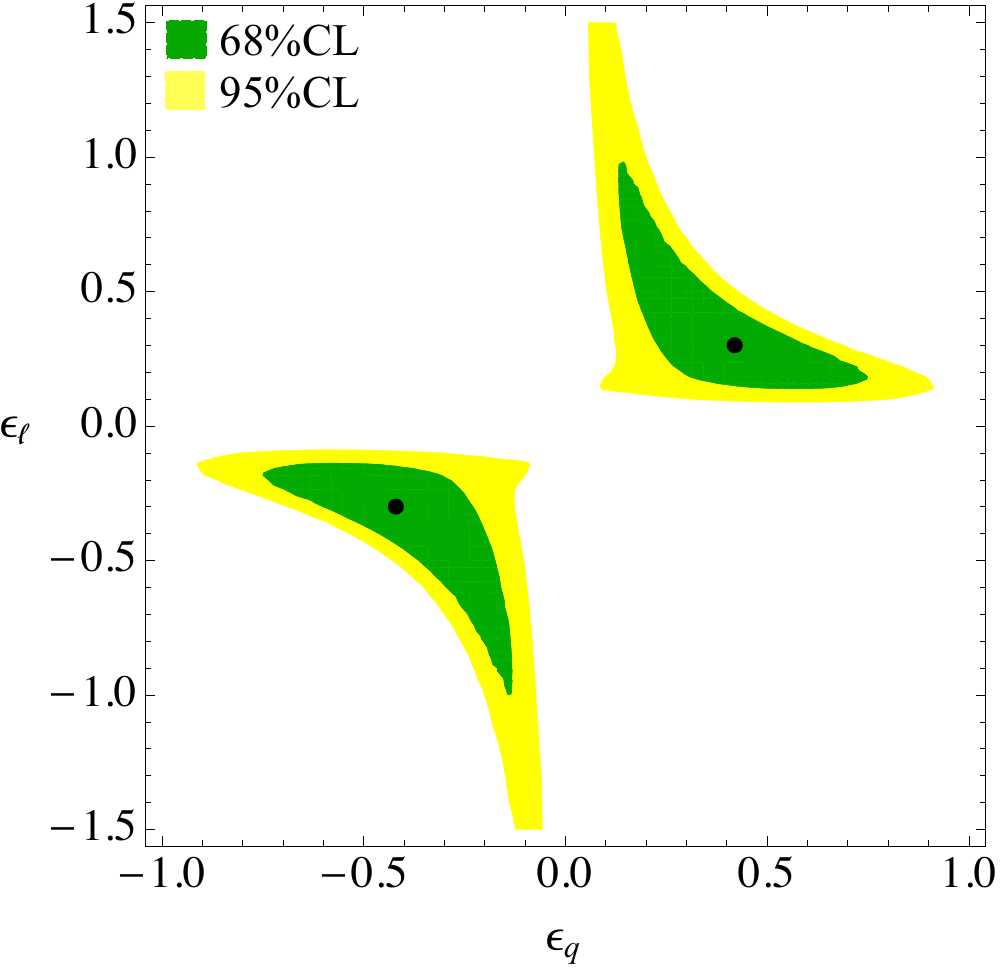} &
   \includegraphics[width=0.47\textwidth]{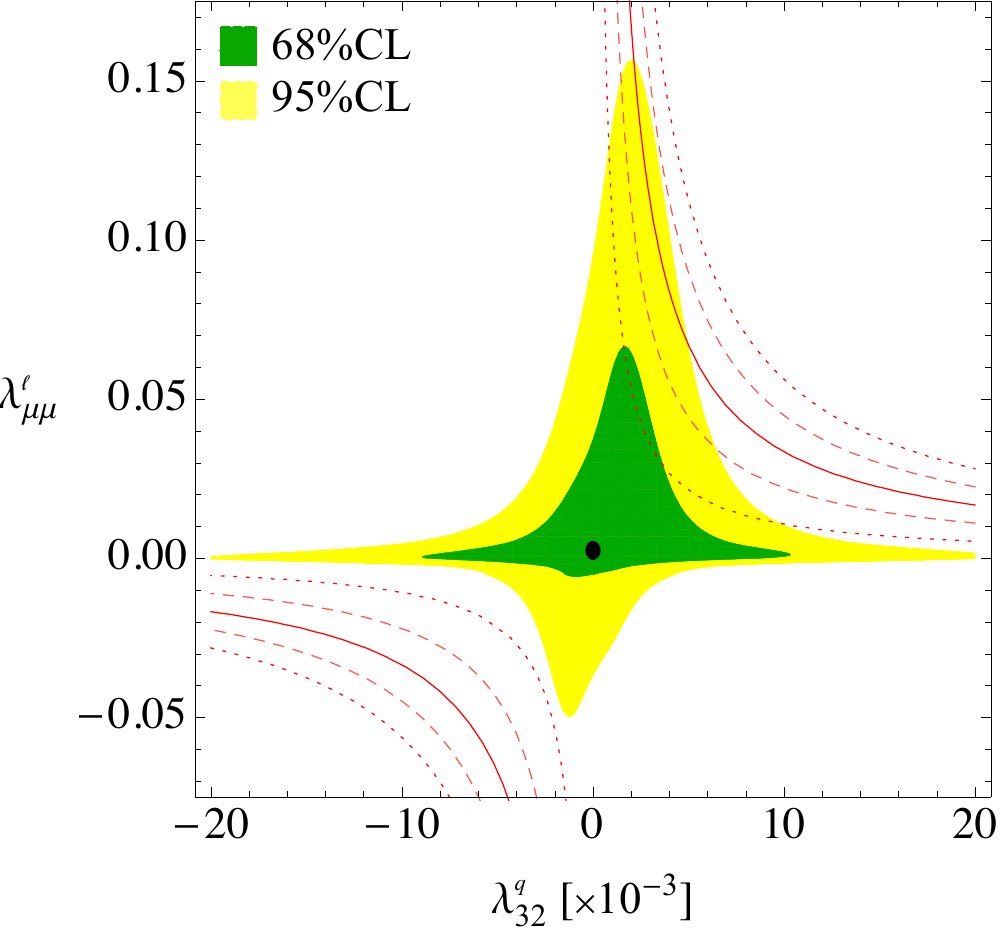}  \\ [0.5 cm]
   \includegraphics[width=0.465\textwidth]{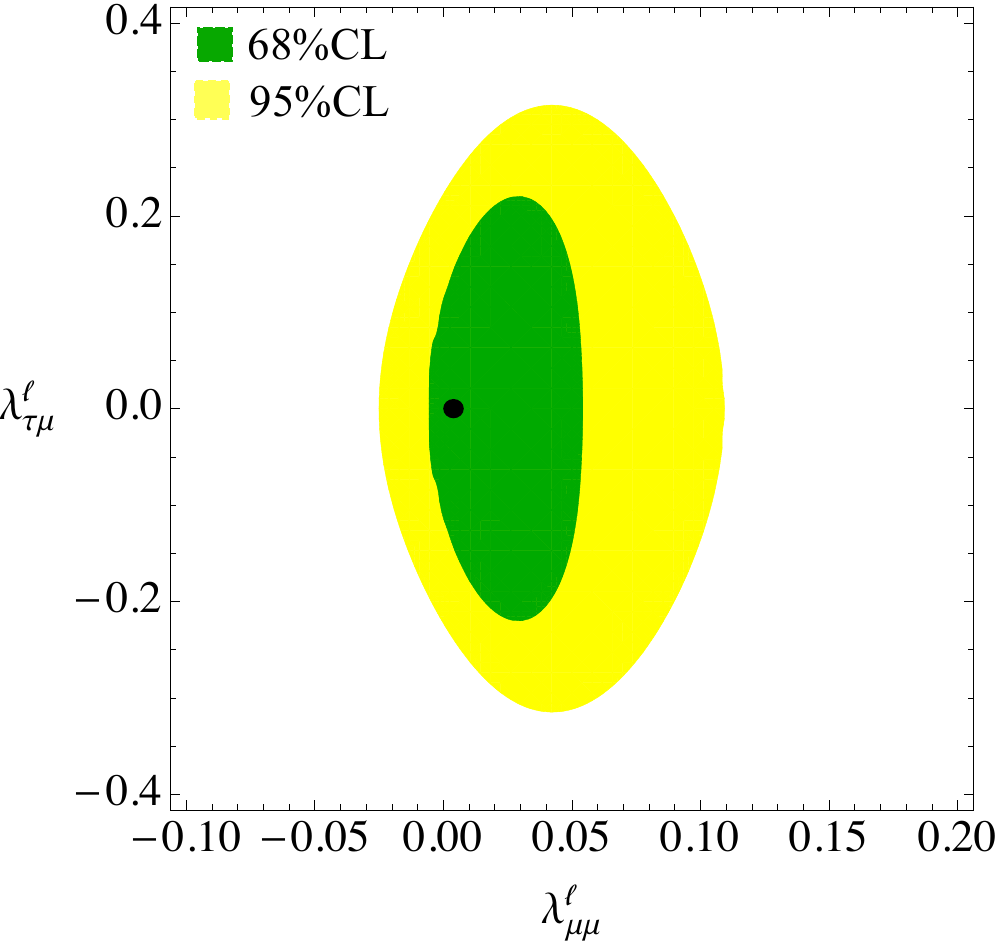} &
   \includegraphics[width=0.475\textwidth]{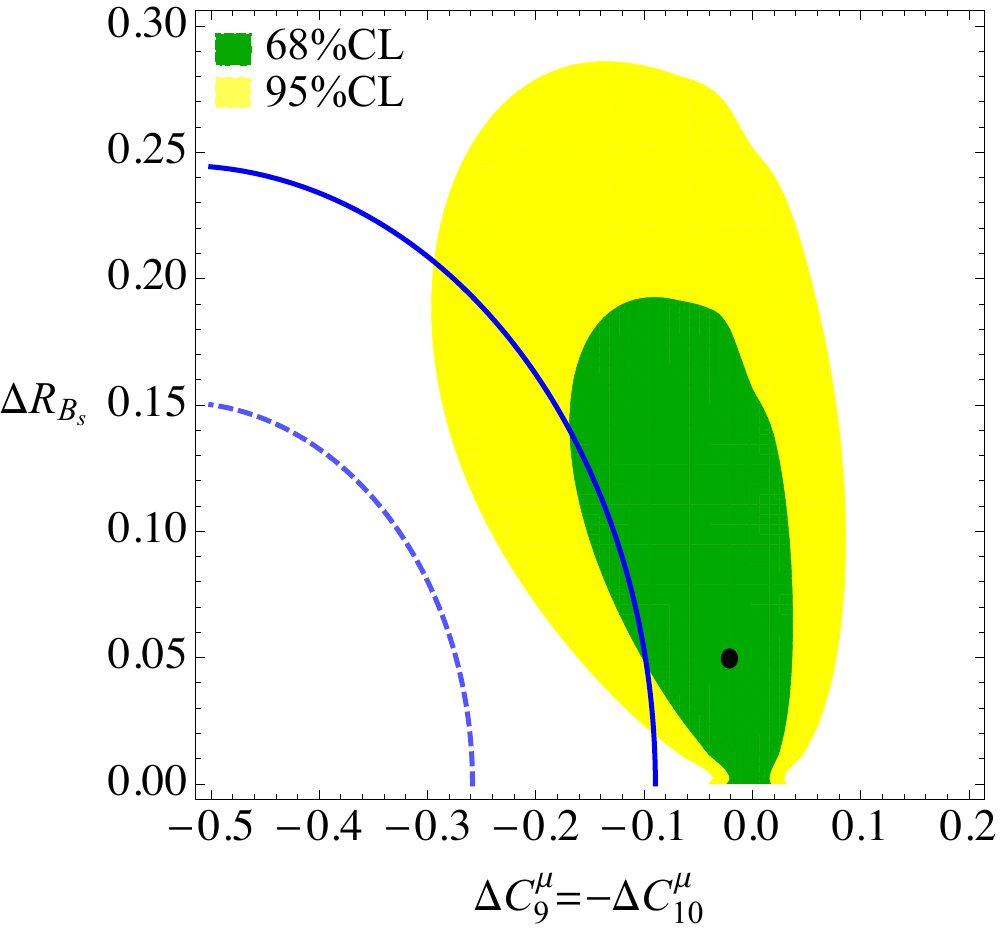}
    \end{tabular}
    \caption{ \small  Results of the  low-energy fit in Table~\ref{tab:FlavorFit}: 68\%CL (green) and 95\%CL (yellow) allowed regions in the $(\epsilon_q, \epsilon_\ell)$ plane (upper-left plot), $(\llq_{bs}, \lle_{\mu\mu})$ plane (upper-right plot), $(\lle_{\mu\mu},\lle_{\tau\mu})$ plane (lower-left plot), and in the $(\Delta C_9^\mu, \Delta R_{B_s})$ plane (lower-right plot), after having marginalised over the variables not shown. The black dots represent the best-fit points for these 2d likelihoods. In the upper-right plot, the solid, dashed, and dotted red lines represent the iso-lines respectively for the best-fit, 1- and 2-$\sigma$ ranges for $\Delta C_9^\mu$, with fixed $R_0 = 0.15$. In the lower-right plot, the dashed and solid blue lines represent the $68\%$CL and $98\%$CL regions for $\Delta C_9^\mu$ and $\Delta  R_{B_s}$ as favored by $b\rightarrow s \mu^+\mu^-$ and $\Delta m_{B_s}$ data.
    \label{Fig:flavor_bounds} }
\end{figure}

The low-energy observables discussed above depend on the three flavor-non-universal couplings 
$\llq_{bs}$, $\lle_{\mu\mu}$, $\lle_{\tau\mu}$, and the two flavor-independent combinations 
\be
	\epsilon_{\ell,q} \equiv \frac{g_{\ell,q} \, m_W}{g \, m_V} \approx g_{\ell,q} \frac{122 \, \GeV}{m_V}~,
\ee
which we assume to be bounded by $|\epsilon_{\ell,q}| < 2$.
We have performed a combined fit of these parameters using the experimental constraints reported in Table~\ref{tab:FlavorFit}.
For simplicity, we have assumed Gaussian errors for all the observables. The preferred region of the model parameters ($x_\alpha$)
has been determined  minimizing the $\chi^2$ distribution
\be
	\chi^2(x_\alpha) = \sum_i \frac{(\mathcal{O}_i(x_\alpha) - \mu_i)^2}{\sigma_i^2}~.
\ee
The best-fit point is found for 
\be
	\epsilon_\ell \approx 0.37~, \quad
	\epsilon_q \approx 0.38~, \quad
	\llq_{bs} \approx 2.3 \times 10^{-3}~, \quad
	\lle_{\mu\mu} \approx 2.0 \times 10^{-2}~, \quad
	\lle_{\tau\mu} \approx 4.8 \times 10^{-2}~.
\ee
The $\chi^2$ improvement of the best-fit point with respect to the SM limit is $\chi^2(x_{\rm SM}) - \chi^2(x_{\rm BF}) =18.6$ for 5 d.o.f., which corresponds to a $p$-value for the SM hypothesis of $0.002$.
In Fig.~\ref{Fig:flavor_bounds} we show the $68\%$CL and $95\%$CL regions in the $(\epsilon_q, \epsilon_\ell)$, $(\llq_{bs},\lle_{\mu\mu})$, $(\lle_{\mu\mu},\lle_{\tau\mu})$, and $(\Delta C_9^\mu, \Delta R_{B_s})$ planes, after having marginalised over the other parameters.

The best-fit point implies a small non-standard contribution to $C^{\mu}_{9}$. This is because of the bounds on $|\lle_{\mu\mu}|$  and $|\llq_{bs}|$ dictated by 
LFU in $\tau$ decays and $\Delta m_{B_q}$ mixing (see sect.~\ref{sect:FCNCs}). However,  in the $95\%$CL ($68\%$CL) preferred region of the  model parameters the effective coupling $|\lle_{\mu\mu}|$ can exceed  $0.10~(0.05)$. 
In this case $\Delta C^{\mu}_{9}$ can be within $1\sigma$ or $2\sigma$ of its  central value (see right panels in 
Fig.~\ref{Fig:flavor_bounds}).\footnote{A ``perfect fit" of  $\Delta C^{\mu}_{9}$ can be obtained extending the minimal version of the model, 
at the cost of introducing more free parameters. In particular, a natural extension is obtained with the inclusion of a $SU(2)_L$ singlet, coupled to the current $J^0_\mu$ obtained by $J^a_\mu$ in Eq.~(\ref{eq:LVf}) with 
the replacement $T^a \to 1, g_{q\ell} \to  g^\prime_{q\ell}$. }

\medskip

In summary, we find that the effective Lagrangian in Eq.~(\ref{eq:DeltaL4f}) provides a significantly improved fit to 
low-energy data. It is worth  stressing that, even from the EFT  point of view, this model is highly constrained
given the  underlying set of dynamical hypotheses. As a result, the model
leads to a  series of predictions 
often different (more precise) than  those obtained using more general 
EFT approaches (see e.g.~\cite{Glashow:2014iga,Bhattacharya:2014wla,Alonso:2015sja}).
The main predictions, which can be used to test the model in a more stringent way in view of future data,
can be summarized as follows: 
\begin{description}
\item[Charged currents.] The $b\to c(u) \tau\nu$ charged currents should exhibit a universal enhancement (independent of the hadronic final state).
This implies, in particular, $R_{B\tau\nu}=R^{\tau/\mu}_{D}= R^{\tau/\mu}_{D^*}$.
 LFU violations  between $b\to c(u) \mu \nu$ and $b\to c(u) e \nu$ can be as large as $O(1\%)$.
 The  inclusive $|V_{cb}|$ and $|V_{ub}|$ determinations are enhanced over the exclusive ones 
 because of the $\tau$  contamination in the corresponding samples.
 \item[FCNC.] The modification of the $b\to s \ell^+\ell^-$ operators are purely left-handed. This implies, in particular,
 $\Delta C_9^\mu = - \Delta C_{10}^\mu$, hence a suppression (in the 
$10\%$--$20\%$ range)  of $\cB(B_{s,d}\to \mu^+\mu^-)$ rates 
 compared to their SM expectations. The central value of the anomaly in $R^{\mu/e}_K$  is likely to decrease 
 (to $\sim 10\%$ or less).
The NP contribution to the $b\to s \tau^+\tau^-$ amplitude is likely to be close, in magnitude, to the SM one. 
This implies a rate enhancement of at most a factor of $\approx 5$ compared to the SM (constructive interference) or a strong suppression
(destructive interference). The magnitude of the FCNC-LFV transitions $b\to s \mu^\pm \tau^\mp$ is at most $10\%$ (in the rates) 
compared to the $b\to s \ell^+\ell^-$ ones. 
\item[Meson-antimeson mixing.] A $O(10\%)$ deviation from the SM is expected $B_s$ mixing, if the 
anomaly in $R^{\mu/e}_K$ persists.  According to the most plausible 
breaking structure of the $U(2)_q$ symmetry, this deviation should be present also in  $B_d$ mixing and should 
preserve the relation  $\Delta M_{B_s}/ \Delta M_{B_d}= (\Delta M_{B_s}/ \Delta M_{B_d})_{\rm SM}$.
 The $D$--$\bar D$ mixing amplitude should acquire a CP-violating phase,  
 whose magnitude could be just below the current experimental bounds.
\item[$\tau$ decays.] The $\tau \to 3\mu$ and $\mu \to 3e$ processes are generated at the tree level (contrary to LFV dipole transitions
$\ell_i \to \ell_j \gamma$) and could be close to the present experimental bounds, although  no precise correlations with other observables can be derived at present.  If the anomaly in $R^{\mu/e}_K$ persists, 
violations of LFU in $\tau \to \mu \bar\nu \nu$ vs.~$\tau \to e \bar\nu \nu$ are expected to be just below the current experimental bounds.
\end{description}

\section{Constraints on the dynamical model}

\subsection{Bounds from LEP-I}
\label{sect:LEP}

Since the couplings of the heavy vector with SM fermions in this model are strongly non-universal, we cannot apply the LEP-I constraints as encoded in the bound on the $S$-parameter. Instead, we consider the non-universal fit of LEP-I data performed in the context of dimension-6 operators in Ref.~\cite{Efrati:2015eaa}. To do this, we translate the effective Lagrangian of Eq.~\eqref{eq:EffLagr} to the \emph{Higgs basis} used for the fit.

Since the constraints on the $Z$ couplings to third generation (left-handed) quarks and leptons are of the same order as the bounds on the couplings to lighter fermions, while in our model the third generation is the one with biggest couplings, the strongest constraints on the model will arise from the bounds on $Z$ couplings to third generation fermions. This motivates us to simplify the analysis of the 
EFT fit by neglecting $\lambda^{q,\ell}_{ij}$ for $i,j \neq 3$.
In this limit the fit only depends on these two combinations of parameters:
\begin{figure*}[t]
  \centering
   \includegraphics[width=0.475\textwidth]{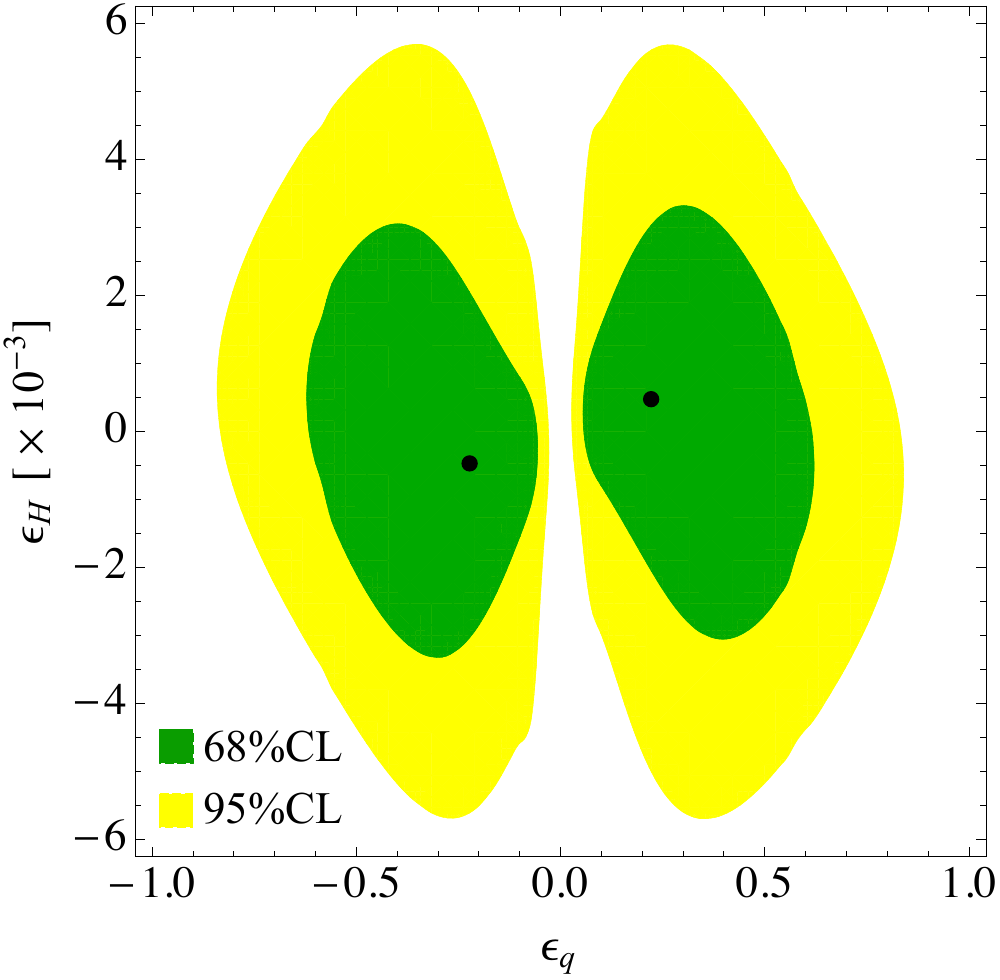}
    \caption{ \small Results of the combined flavor and electroweak fit: 68\%CL (green) and 95\%CL (yellow) allowed regions in the $(\epsilon_q, \epsilon_H)$ plane after having marginalized over the other parameters. The black dots represent the best-fit points. \label{Fig:B_LEP1_bounds}}
\end{figure*}
\be
	\epsilon_\ell \, \epsilon_H \equiv \frac{\gl g_H m_W^2}{g^2 m_V^2} = (4.3 \pm 8.7) \times 10^{-4}~, \qquad
	\epsilon_q \, \epsilon_H \equiv \frac{\gq g_H m_W^2}{g^2 m_V^2} = (- 0.8 \pm 1.4) \times 10^{-3}~,
\ee
and the correlation is negligible. We introduced the adimensional parameters $\epsilon_X \equiv g_X m_W / g m_V$, with $X = \ell, q, H$.
With this notation the constraint in Eq.~\eqref{eq:BoundR0} from charged current $B$-decays can be written as $R_0 = \epsilon_\ell \, \epsilon_q = 0.14 \pm 0.04$.
 In Fig.~\ref{Fig:B_LEP1_bounds} we combine these experimental constraints with the ones from flavor physics and show the $68\%$CL and $95\%$CL allowed regions in the $(\epsilon_q, \epsilon_H)$ plane. From this we conclude that $|\epsilon_H| \lesssim 5 \times 10^{-3}$. 
This result allows us  to conclude that, in absence of new degrees of freedom in the model, the massive vectors decay dominantly to SM 
fermions by means of the last term ($V^a_\mu J_\mu^a$) in Eq.~(\ref{eq:dyn_mod}).
	
\subsection{High-energy searches} 
 
We parametrize massive vector boson  couplings to SM fermions (in their mass-eigenstate basis) as follows 
\be
\Delta \cL_ {VJ}  =   V^a_\mu J_\mu^a = c^V_{i j} ~\bar f^i_L \gamma^\mu f^j_L V_\mu~.
\label{eq-Vqq}
\ee
With this definition, the two body $V \to \bar f_i f_j$ decay width is
\be
\Gamma(V\to \bar f_i f_j) = \frac{m_V}{24\pi} N_C | c^V_{i j}|^2 \mathcal{F}\left(\frac{m_{f_i}}{m_V},\frac{m_{f_j}}{m_V}\right)~,
\label{eq-tot-DW}
\ee
 where
 \be
 \mathcal{F}(x,y)=\left(1-\frac{x^2+y^2}{2}-\frac{(x^2-y^2)^2}{2}\right)\sqrt{1-2(x^2+y^2)-(x^2-y^2)^2}~,
 \ee
$N_C$ is the dimension of the color representation of the fermions, and we have assumed 
$m_V > m_{f_i}+m_{f_j}$. 
 
Due to the approximate $U(2)_q \times U(2)_\ell$ symmetry, the total decay width of the vector bosons is dominated by
decays to third-generation fermions. In the limit  $m^2_V \gg 4 m^2_t$, 
\be
\frac{\Gamma_{V^\pm}}{m_{V^\pm}}\approx \frac{\Gamma_{V^0}}{m_{V^0}} \approx  \frac{1}{48\pi}(g_\ell^2+3 g_q^2)~.
\ee
The neutral vector boson predominantly decays to $\tau^+ \tau^-$, $\bar\nu_\tau \nu_\tau$, $\bar{b} b$, and $\bar t t$ final states.
The relative impact of the leptonic and hadronic decay modes is driven by the ratio $g_\ell/g_q$ (and the phase-space corrections to the large top-quark mass). The decay to a muon pair is parametrically suppressed by the smallness of $\lle_{\mu\mu}$. In particular, the following relation holds 
\be
\mathcal{B}(V^0\to\mu^+\mu^-)=|\lle_{\mu\mu}|^2~ \mathcal{B}(V^0\to\tau^+\tau^-)~.
\ee
The dominant charged vector decay modes are $t \bar b$ and $\tau^+ \nu_\tau$. In the following we assume $m_{V^+} > m_t$, such that the $V^+$ state cannot be produced on-shell from top decays. We checked by explicit computation that when this criteria is satisfied the corrections to $t\to b \tau^+ \nu_\tau$ decay are well below present experimental sensitivity. The decays of both charged and neutral states to 
SM gauge bosons are strongly suppressed due to the strong limits on the $\epsilon_H$ parameter from electroweak precision data
(see Sect.~\ref{sect:LEP}).

The single vector bosons production at the LHC is dominated by Drell-Yan type processes, i.e.~$p p \to V+X$, where $X$ stands for additional hadronic activity. While resonance searches in general impose severe limits on sequential (SM--like) $W'$ and $Z'$ bosons, we find significantly milder limits within our model. This is because of the specific flavor structure which suppresses both the production cross section and the decays into muon 
or electron final states.

In order to derive the present collider limits on the model, we have confronted the predictions of the model to a number of ATLAS and CMS searches for heavy $W'$ and $Z'$ resonances~\cite{Aad:2014xea,CMS:2015vda,ATLAS:2014wra,Aad:2015osa,Aad:2014cka,Khachatryan:2015sja,Chatrchyan:2012ku}. To this purpose, we have implemented the model in Universal FeynRules Output (UFO)~\cite{Degrande:2011ua} using the FeynRules~\cite{Christensen:2008py} package version 2.3.1. We have used  the MG5\_aMC\_v2.2.3~\cite{Alwall:2014hca} package to simulate the tree-level $p p \to V^{\pm}$ and $p p \to V^0$ production at $\sqrt{s}=8$ TeV in the 5-flavor scheme.  Finally, we have validated the implementation of the heavy vector triplet couplings to fermions by simulating decays and comparing the numerical results with the analytic
expressions in Eq.~(\ref{eq-tot-DW}).

 \begin{figure}[p]
  \begin{center}
    \includegraphics[width=0.505\textwidth]{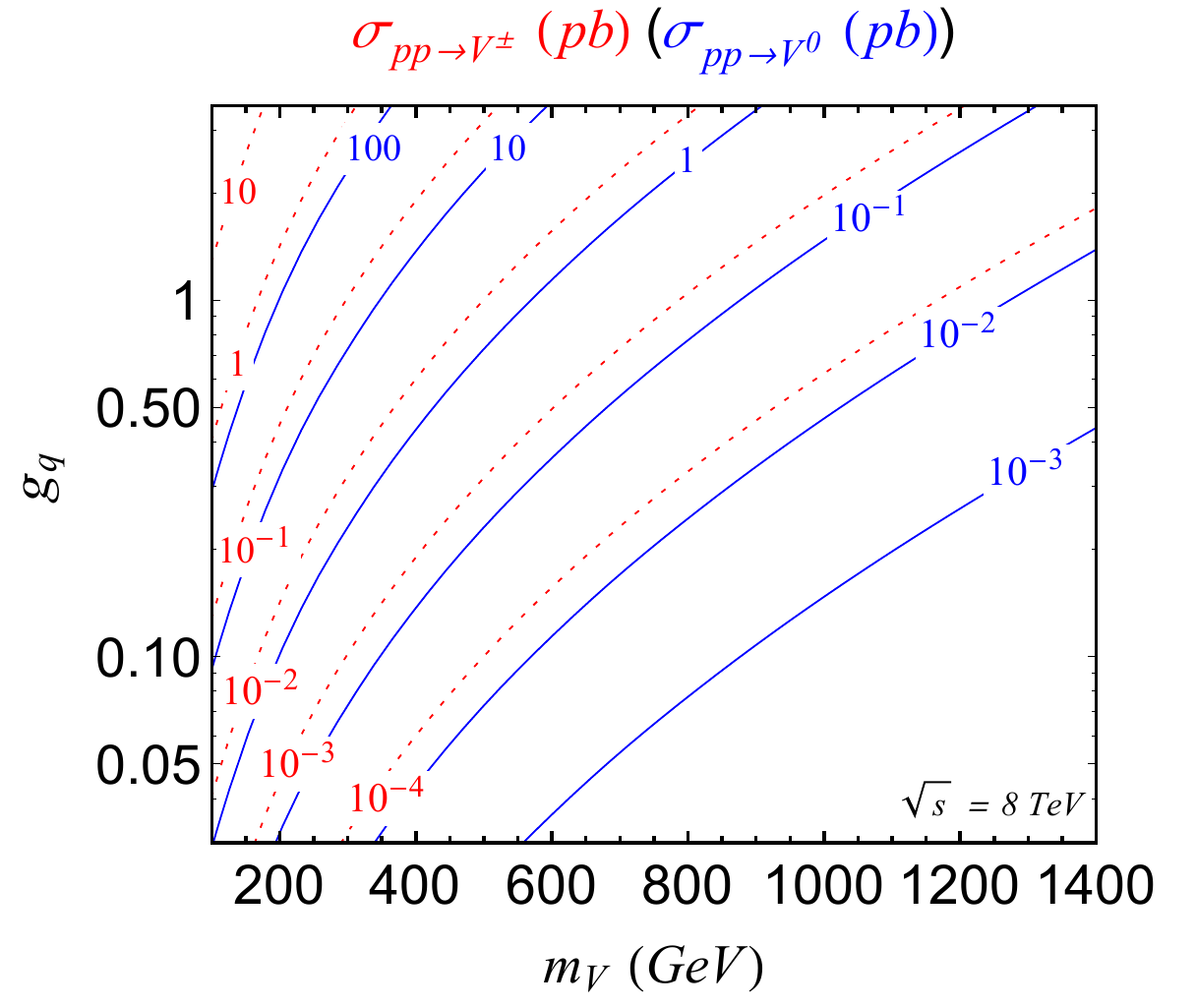}
     \includegraphics[width=0.485\textwidth]{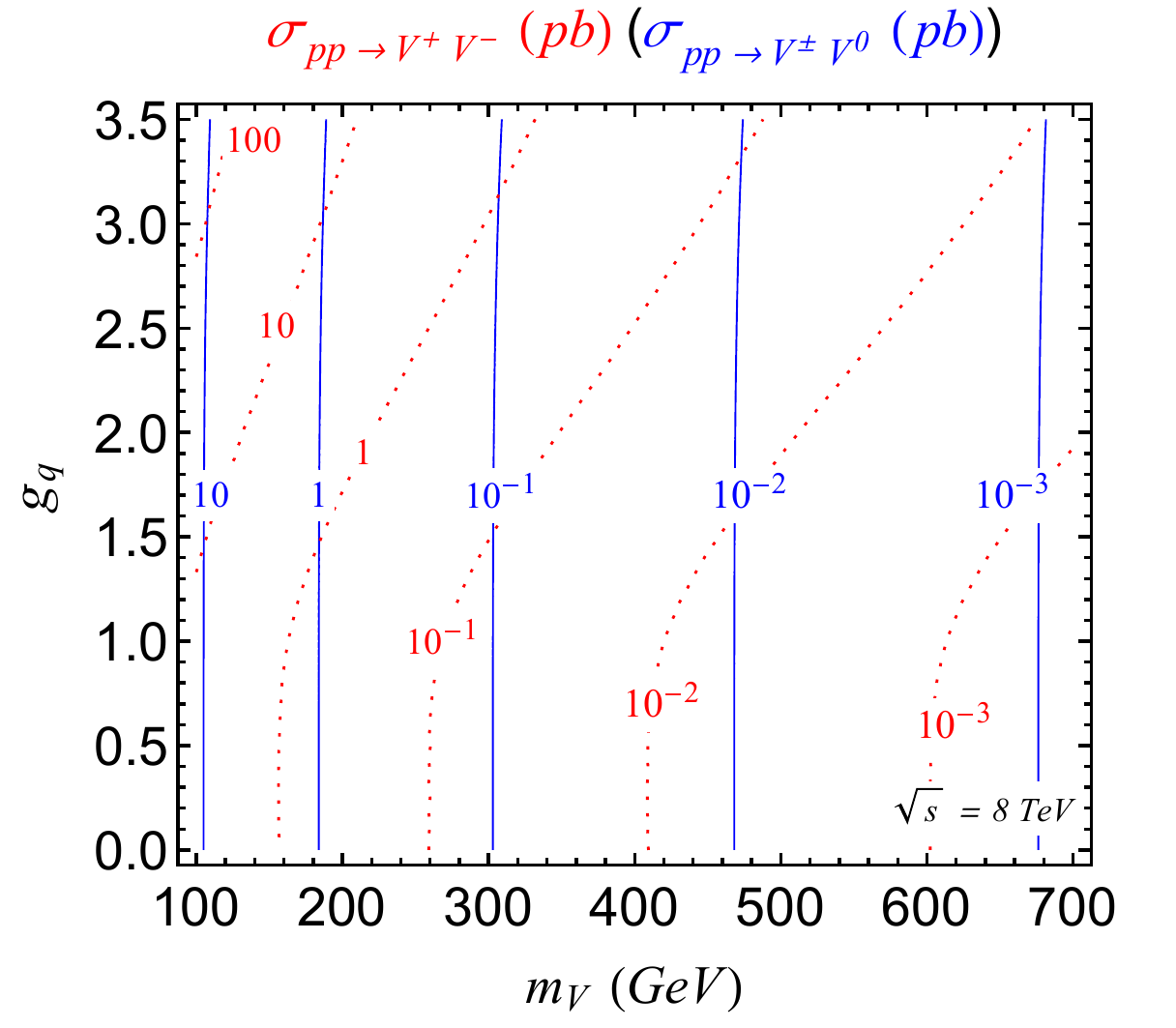}
  \end{center}
\caption{\label{fig-2} Predicted cross sections for single and pair production of charged and neutral vector bosons for $8$~TeV $pp$ collisions as a function of $g_q$ and $m_V$ in the limit of approximate flavor symmetry.}
\end{figure}

In Fig.~\ref{fig-2}~(left) we show the  predicted cross sections (in pb) for  $p p \to V^{\pm}$ ($p p \to V^{+}$ plus $p p \to V^{-}$) and $p p \to V^{0}$,
as obtained in the limit $\llq_{ij} = \delta_{i3}\delta_{3j}$ (i.e.~exact flavor symmetry, but for the breaking terms 
induced by the SM Yukawa couplings).  In this limit the production cross sections are 
 completely determined by $g_q$ and $m_V$. As can be seen, the neutral cross-section is about 100 times larger than the charged one.
 This is because the $V^0$ state is produced by bottom-bottom fusion, that is allowed in the limit of exact flavor symmetry,
 while the leading $V^{\pm}$ production channel is bottom-charm fusion, that is suppressed by $|V_{cb}|$
 at the amplitude level. 
   
 The search for $W' \to t \bar b$ with the ATLAS detector at $8$~TeV and $20.3$~fb$^{-1}$ of data excludes a left-handed $W'$ boson with a mass of $500~(1000)$~GeV if $\sigma(p p \to W')\times \mathcal{B}(W'\to t \bar b) > 3.3~(0.19)$~pb~\cite{Aad:2014xea}. In addition, the CMS search for $W' \to \tau \nu$ performed at $8$~TeV with an integrated luminosity of $19.7$~fb$^{-1}$ excludes a $W'$ boson with mass $300~(500)$~GeV if $\sigma(p p \to W')\times \mathcal{B}(W'\to \tau \nu) > 4~(0.1)$~pb~\cite{CMS:2015vda}. Comparing these limits with the predicted cross sections shown in 
 Fig.~\ref{fig-2}~(left), we conclude that these searches have little impact on our model. We also checked that the ATLAS search for $W'\to \mu \nu$~\cite{ATLAS:2014wra} has no relevance due to the limit on the $|\lle_{\mu\mu}|$ coupling.

\begin{figure}[p]
  \begin{center}
     \includegraphics[width=0.59\textwidth]{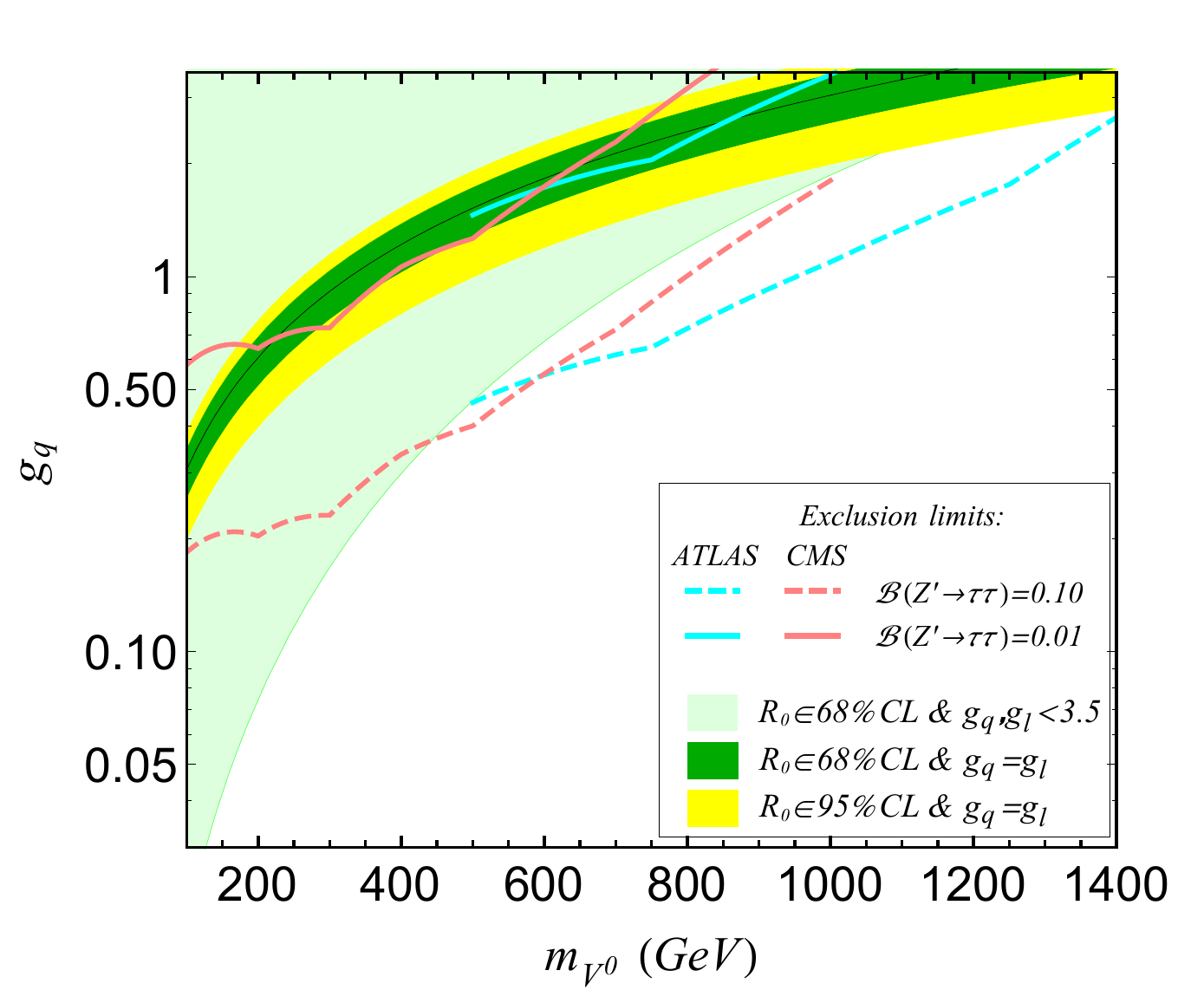}
  \end{center}
\caption{\label{fig-3} Preferred region from flavor data and exclusion limits from LHC. See text for details.}
\end{figure}

The resonance searches for neutral vector bosons are more relevant due to the larger expected cross section. 
The ATLAS search for $Z'$ resonances decaying to $\tau^+ \tau^-$ using $19.5-20.3$~fb$^{-1}$ of 8~TeV data~\cite{Aad:2015osa} sets an important constraint on the parameter space of the model. The exclusion limits (under the assumption of a narrow resonance) are shown in 
Fig.~\ref{fig-3} in cyan solid (dashed) line assuming $\mathcal{B}(Z'\to \tau^+ \tau^-)=0.01~(0.10)$. The region above these lines is excluded  at 95\%CL. The exclusion limits start from $m_V=500$~GeV because Ref.~\cite{Aad:2015osa} reports the limits on $\sigma\times\mathcal{B}$ only 
above   this mass. To overcome this problem, we urge the experimental collaborations to extend the search for $Z'$ resonances even in the low mass region. To extract the present limits  for $m_V< 500$~GeV, we use the CMS search for the neutral MSSM Higgs boson decaying to a pair of tau leptons~\cite{Khachatryan:2014wca} at $8$~TeV and $19.7$~fb$^{-1}$ luminosity. The collaboration reports a model independent limit on the $b \bar{b}$--induced   production cross section times $\mathcal{B}(H\to\tau^+\tau^-)$ (assuming a narrow resonance) in the region 
$100~{\rm GeV} < m_H < 1$~TeV. We have performed a parton-level MadGraph simulation to compare the kinematics of the $\tau^+\tau^-$ 
final state produced by a scalar and a vector resonance (of mass 200~GeV). Having found small differences, we have re-interpreted the CMS 
bound into the $\sigma\times\mathcal{B}$ limit for our model reported in Fig.~\ref{fig-3}.  In particular, the region above pink solid (dashed) line is excluded assuming $\mathcal{B}(Z'\to \tau^+ \tau^-)=0.01~(0.10)$. While the searches for dimuon resonances are usually more sensitive than the $\tau^+ \tau^-$ ones (see for instance~\cite{Aad:2014cka}), we find them less relevant for our model  due to smallness of $\lle_{\mu\mu}$. On the other hand, dijet~\cite{Khachatryan:2015sja} and $t\bar t$~\cite{Chatrchyan:2012ku} resonance searches set limits on the cross section times branching ratio for  $m_V \sim 1$~TeV of the order of $1$~pb.

The impact of the direct searches in the $\tau^+\tau^-$ channel on the parameter region preferred by flavor data is illustrated in Fig.~\ref{fig-3}. The wide light green region is obtained imposing $R_0=0.14\pm0.04$ ($68\%$~CL region) and $g_q,g_\ell<\sqrt{4\pi}$. 
The narrower dark green (yellow) band is the region for which $R_0$ is within $68\%$~CL ($95\%$~CL) and $g_q =g_\ell$. In the minimal model, the predicted $V^0\to \tau^+ \tau^-$ branching ratio for $g_q =g_\ell$ is  $\mathcal{B}( V^0\to \tau^+ \tau^- ) \approx 1/8$. Comparing with the exclusion
curve obtained for $\mathcal{B}(Z'\to \tau^+ \tau^-)=0.1$ we deduce that the minimal model is ruled-out for $g_q =g_\ell$.  The situation improves 
for $ g_\ell \ll g_q$ (a configuration also preferred by flavor data, albeit with the lower bound $g_\ell \gtrsim g_q /5.4$ from $D$--$\bar{D}$ mixing, 
see sect.~\ref{sect:DF2}), given $\mathcal{B}( V^0\to \tau^+ \tau^- ) \approx  (1/8) \times (g_\ell / g_q)^2$.
However, it is not possible to completely evade the bounds for perturbative values of the couplings ($g_q,g_\ell<\sqrt{4\pi}$) in the region  preferred
by flavor data.

There are different ways to evade the LHC limits on $pp \to V^0 \to \tau^+ \tau^-$ going beyond the minimal model.
The simplest possibility is to add new $V^0$ decay channels, say to a dark sector. This would result into  lower values of $\mathcal{B}( V^0\to \tau^+ \tau^- )$. As can be seen  in Fig.~\ref{fig-3}  if, for $g_q =g_\ell$,  $\mathcal{B}( V^0\to \tau^+ \tau^- )$ decreases to $0.01$, then   there 
are regions of the parameter space that are allowed, both at low and at high masses.  Another option is to consider a heavy $V^0$ in the limit of a 
strongly coupled theory ($\Gamma_V \sim M_V$). In this case the  resonance becomes broad and the limits obtained assuming a narrow state 
no longer holds. A third possibility would be to add an additional neutral heavy vector, e.g.~a $SU(2)_L$ singlet, close in mass to the 
neutral component of the triplet, with couplings tuned to interfere destructively with $V^0$ in the $pp\to \tau^+\tau^- +X $ cross section.

We finally comment about the $p p \to V V$ (pair production) process at the LHC. This proceeds via: i)~$t$-channel quark-exchange diagrams, controlled by 
Eq.~(\ref{eq-Vqq}), and ii)~$s$-channel diagrams with off-shell SM electroweak gauge bosons. In the limit of no mixing with EW gauge bosons, $g_H \approx 0$, and neglecting contributions from additional non-minimal operators \cite{Pappadopulo:2014qza}, 
the relevant interactions of the heavy vectors with the SM electroweak gauge bosons described by Eq.~\eqref{eq:dyn_mod} are
\ba
\mathcal{L} &\supset& - i g (s_\theta A_\mu + c_\theta Z_\mu) V^-_\nu (\partial^\mu V^{+\nu}-\partial^\nu V^{+\mu}) - i g W^{+\mu} V^{0\nu} (\partial_\mu V^{-}_{\nu} -\partial_\nu V^{-}_{\mu})~ \no\\
&&  + i g W^+_\mu V^-_\nu ( \partial^\mu V^{0\nu}-\partial^\nu V^{0\mu})+\rm{h.c.}~,
\ea
where $g$ is the $SU(2)_L$ coupling constant and $s_\theta$ ($c_\theta$) is sine (cosine) of the Weinberg angle.
As for the single $V$ production, we have used MadGraph to simulate $p p \to V^+ V^-$ and $p p \to V^\pm V^0$ (that is, $p p \to V^+ V^0$ plus $p p \to V^- V^0$). In Fig.~\ref{fig-2}~(right) we show the predicted cross sections (in $pb$) for 8 TeV proton-proton collisions, as obtained in the limit $\llq_{ij} = \delta_{i3}\delta_{3j}$. As illustrated with the vertical iso-lines, the cross sections are dominated by $s$-channel diagrams for small $g_q$ couplings. On the other hand, for large couplings there is a substantial contribution from the $b \bar b$ induced $t$-channel diagram to $V^+ V^-$ production. Based on the predicted cross sections, we conclude that the single production is more relevant compared to the pair production for the interesting region of the parameter space.

\section{Conclusions}

Lepton Flavor Universality is not a fundamental symmetry: within the Standard Model it is an approximate accidental symmetry 
broken only by the Yukawa interactions. This specific symmetry and symmetry-breaking pattern results in tiny deviations 
from LFU in helicity-conserving amplitudes, within the SM, and it implies that LFU tests are clean probes of physics beyond the SM.

Motivated by a series of recent experimental results  in $B$ physics pointing to possible violations of LFU, both in charged and in neutral currents, in this paper we have consider a simplified dynamical model able to describe these effects in a unified way. In particular, we have 
shown that a $SU(2)_L$ triplet of massive vector bosons, coupled predominantly to third generation fermions 
(both quarks and leptons),  can significantly improve the description of present data.

The proposed model has a series of virtues compared to previous attempts to describe such effects 
in terms of New Physics: i) it connects the breaking of LFU between charged and neutral currents, 
and between semi-leptonic and purely leptonic processes; 
ii) it is based on a simple flavor symmetry, whose breaking terms are related to the structure of the SM Yukawa 
couplings, both in the quark and in the lepton sector; iii) it connects low-energy deviations from the SM to 
direct searches for NP at high $p_T$. The constrained structure of the model makes it highly non trivial to satisfy 
all existing bounds and, at the same time, accommodate deviations from the SM as large as 
indicated by the central values in Eqs.~(\ref{eq:RDexp})--(\ref{eq:RKexp}). We find that this happens quite 
naturally in the case of charged currents, both at low and at high energies. The situation is more problematic 
in the case of neutral currents. On the one hand, the maximal deviations from unity in $R^{\mu/e}_K$ barely exceed
$10\%$. On the other hand, the minimal version of the model is ruled out by the direct searches for resonances 
decaying into $\tau^+\tau^-$ at ATLAS and CMS. As discussed, both these issues can be improved with 
less minimal versions of the model, at the cost of introducing more free parameters. 

One of the most remarkable aspects of the minimal version of the model is the well-defined pattern of deviations 
in low-energy processes listed at the end of Sect.~\ref{sect:fit}. This pattern is largely insensitive to possible extensions of the 
model necessary to overcome the constraints from direct searches. It mainly reflects the symmetry structure of the 
model and could be used, in the near future, to verify or falsify this 
framework  with more precise data. 

Besides the specific predictions of the proposed model, we stress the importance of future experimental tests 
of LFU at low-energies, and dedicated searches for flavor-non-universal phenomena at high energies. 
On the low-energy side, we urge the experimental collaborations to re-analyze charged-current $B$ decays without assuming 
lepton flavor universality. On the high-energy side, we encourage the search for deviations from the SM in 
$\tau^+\tau^-$ and $t\bar t$ invariant-mass distributions, relaxing the hypothesis of narrow resonances and covering 
also the region of low invariant masses.

\section*{Acknowledgements}
We thank Riccardo Barbieri, Andreas Crivellin, Martin Schmaltz, and Nicola Serra for useful  comments and discussions.
This research was supported in
part by the Swiss National Science Foundation (SNF) under contract 200021-159720.

\end{document}